\documentclass[twocolumn,aps,prl,showpacs,amsmath,superscriptaddress,longbibliography,notitlepage]{revtex4-1}
\usepackage{amssymb}
\usepackage{mathrsfs}
\usepackage{graphicx}
\usepackage{float}
\usepackage[caption=false]{subfig}
\usepackage[pdftex,colorlinks=red]{hyperref}

\usepackage{epstopdf}

\usepackage[normalem]{ulem}
\usepackage{verbatim}
\usepackage{xcolor}
\usepackage{bm}

\def\red{\textcolor{red}}
\def\blue{\textcolor{blue}}
\def\CH{\textcolor{magenta}}

\begin{document}

\title{Non-Hermitian Pseudo-Gaps}

\author{Linhu Li}  \email{lilh56@mail.sysu.edu.cn}
\affiliation{Guangdong Provincial Key Laboratory of Quantum Metrology and Sensing $\&$ School of Physics and Astronomy, Sun Yat-Sen University (Zhuhai Campus), Zhuhai 519082, China}
\author{Ching Hua Lee}  \email{phylch@nus.edu.sg}
\affiliation{Department of Physics, National University of Singapore, Singapore 117551, Republic of Singapore}

\begin{abstract}
The notion of a band gap is ubiquitous in the characterization of matter. Particularly interesting are pseudo-gaps, which are enigmatic regions of very low density of states that have been linked to novel phenomena like high temperature superconductivity. In this work, we discover a new non-Hermitian mechanism that induces pseudo-gaps when boundaries are introduced in a lattice. It generically occurs due to the interference between two or more asymmetric pumping channels, and possess no analog in Hermitian systems. 
Mathematically, it can be visualized as being created by divergences of spectral flow in the complex energy plane, analogous to how sharp edges creates divergent electric fields near an electrical conductor. 
A non-Hermitian pseudo-gap can host symmetry-protected mid-gap modes like ordinary topological gaps, but the mid-gap modes are extended instead of edge-localized, and exhibit extreme sensitivity to symmetry-breaking perturbations.
Surprisingly, pseudo-gaps can also host an integer number of edge modes even though the pseudo-bands possess fractional topological windings, or even no well-defined Chern number at all, in the marginal case of a phase transition point. Challenging conventional notions of topological bulk-boundary correspondences and even the very concept of a band, pseudo-gaps post profound implications that extend to many-body settings, such as fractional Chern insulators. 
\end{abstract}

\date{\today}

\maketitle

\noindent{\it Introduction.--} Band gaps play a central role in intriguing condensed matter phenomena, from metal-insulator transitions to 
superconductivity to topological phases. These phenomena are typically protected by band gaps or quasiparticle excitation gaps, which are induced through physical mechanisms such as cooper paring for superconductors \cite{BCS_paper,BCS_book} and spin-orbit couplings for topological insulators \cite{SO_topo,bernevig2013topological}.  However, band gaps are not always cleanly defined, such as the case of pseudo-gaps, which resemble band gaps but are in reality regions of very low density of states (DOS). Possessing vestiges of both gapped and gapless scenarios, the enigmatic role of pseudogaps in high-temperature superconducting cuprates and non-Fermi liquids have mystified physicists for decades~\cite{emery1995importance,kaminski2002spontaneous,lee2006doping,fauque2006magnetic,xia2008polar,Vojta2009PG,varma2010mind,tahir2011PG,yamaji2011composite,mei2012luttinger,shekhter2013bounding,mishra2014effect,keimer2015PG,proust2016fermi,battisti2017PG,zhang2020pseudogap}.

In this work, we introduce a new type of pseudo-gap arising from a novel non-Hermitian (NH) mechanism. \CH{It} exists because NH systems are special in at least two fundamental ways. Firstly, their spectrum is not constrained to be real, and can thus acquire geometric and topological features in the complex energy plane, such as point-gapped loops without Hermitian analog~\cite{xiong2018does,gong2018topological,zhang2019correspondence,okuma2020topological,li2021quantized,su2020direct,wang2021generating}. 
Secondly, with point gaps, NH lattices also experience the non-Hermitian skin effect (NHSE) marked by dramatic boundary mode accumulation with universal spectral flow in the complex energy plane
~\cite{yao2018edge,yokomizo2019non,Lee2019anatomy,yang2020non,lee2020unraveling}. 

Our work shows how the combination of these two fundamental features create pseudo-gaps with arbitrarily low DOS, going beyond previous theoretical~\cite{xiong2018does,shen2018topological,gong2018topological,li2021quantized,
kawabata2019symmetry,
yao2018edge,yokomizo2019non,
Lee2019anatomy,
kunst2018biorthogonal,Yao2018nonH2D,Yin2018nonHermitian,Hui2018nonH,li2019geometric,song2019realspace,song2019non,okuma2019topological,mu2020emergent,
jiang2019interplay,longhi2019topological,kunst2019non,
Song2019BBC,borgnia2020nonH,
Lee2019hybrid,li2020topological,
okuma2020topological,zhang2019correspondence,wang2020defective,
lee2020unraveling,chang2020entanglement,li2020critical,budich2020sensor,lee2020ultrafast,yang2020non,yi2020nonH,lee2020many,lee2020exceptional,mandal2020nonreciprocal,teo2020topological,okuma2021quantum,li2021impurity,zhang2021tidal,zhang2021universal,yang2021dissipative,song2021non} 
and experimental~\cite{su2020direct,wang2021generating,helbig2020generalized,xiao2020non,ghatak2020observation,weidemann2020efficient,zou2021observation,stegmaier2021topological} works 
where adiabatic continuity between the periodic and open boundary condition (PBC and OBC) spectra is generally assumed. As such, besides reformulating major notions like topological bulk-boundary correspondences and criticality~\cite{xiong2018does,yao2018edge,okuma2020topological,lee2020many,li2020critical}, the NHSE here also raises fundamental questions on the nature of topological band gaps, with implications like quasi-particle fractionalization.

\begin{figure}
\includegraphics[width=1\linewidth]{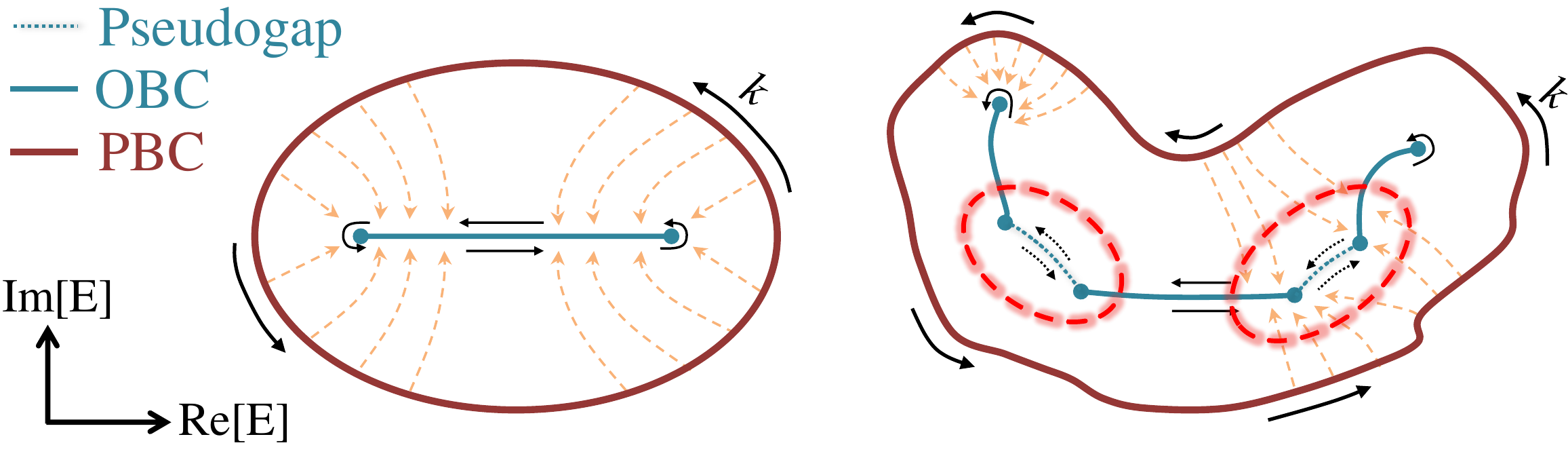}
\caption{Emergence of pseudo-gaps in non-Hermitian OBC spectra. In 1D NH Hamiltonians, PBC spectra (brown) generically trace out closed loops in the complex $E$ plane. 
Their OBC spectra (cyan) are obtained by conformally shrinking PBC loops (dashed orange arrows) until they form lines or curves within their interiors. (Left) Ordinarily, a PBC loop shrinks into one OBC curve, with 1:1 correspondence between OBC and PBC bands. (Right) However, depending on PBC loop curvature, two or more distinct OBC  curves (pseudo-bands) can also form, separated by regions of very low DOS (dotted cyan, circled in red) known as NH pseudo-gaps.
}
\label{fig:PBC-OBC}
\end{figure}

\noindent{\it Origin of the NH pseudo-gap.--} To understand the mechanism behind  NH pseudo-gaps, we consider a 1D NH lattice chain~\footnote{The mechanism for NH pseudo-gaps applies in higher-dimensional lattices, but we specialize to 1D for simplicity.} with PBC spectrum $E(k)$, $k\in [0,2\pi)$ tracing out a loop in complex energy space (brown loops in Fig.~\ref{fig:PBC-OBC}). The loop generically enclose a non-zero area if non-reciprocity exists in the couplings. It is established~\cite{xiong2018does,yao2018edge,Lee2019anatomy} that upon the introduction of spatial inhomogeneities such as open boundaries (OBCs) , the PBC spectral loop will shrink into its interior (via orange arrows), such that it collapses into a line, curve or tree-like structure~\cite{lee2020unraveling} (cyan). This extensive spectral flow moves the eigenvalues far away from the bulk (PBC) spectrum, implying that \emph{all} states must accumulate at a physical boundary. Mathematically, the accumulation is represented by the complex deformation $k\rightarrow k+i\kappa(k)$ with $E(k+i\kappa(k))$ being the degenerate OBC spectrum (cyan in Fig.~\ref{fig:PBC-OBC}), such that an original Bloch state $\psi_k(x)\sim e^{ikx}\rightarrow e^{ikx}e^{-\kappa(k)x}$ acquires an exponential decay term. In most cases (Left in Fig.~\ref{fig:PBC-OBC}), there exists a clear 1-to-1 correspondence between the PBC (outer brown loop) and OBC (internal cyan line) bands. However, it is also possible that the OBC band breaks into multiple pseudo-bands separated by pseudo-gaps (Right, dotted cyan and circled red). These are not true gaps due to adiabatic continuity inherited from the continuous PBC loop. However, they are pseudo-gaps in the sense that their DOS can be made arbitrarily low.

For general insights into when pseudo-gaps occur, we note that as we interpolate from PBCs to OBCs, the shrinking of the spectral loop is controlled by the conformal map $E(k)$, with both $E$ and $k$ regarded as complex variables. As such, the spectral flow lines traced out as $\text{Im}[k]=\kappa$ are varied (orange in Fig.~\ref{fig:PBC-OBC}) are perpendicular to the PBC loop with $\text{Im}[k]=0$ (brown), mathematically analogous to the electric field lines emanating from a conductor~\cite{schinzinger2012conformal}. In particular, they diverge/converge if the PBC loop (``conductor'') is convex/concave on the inside. As such, by engineering the geometry i.e. curvature of the PBC spectrum, one can arrange for the divergences to meet at certain segments of the OBC spectrum (Fig.~\ref{fig:PBC-OBC} Right). As demonstrated in the model below, a pseudo-gap with very low DOS can already be produced even from a benign-looking model with simple couplings.

\begin{figure}
\includegraphics[width=1\linewidth]{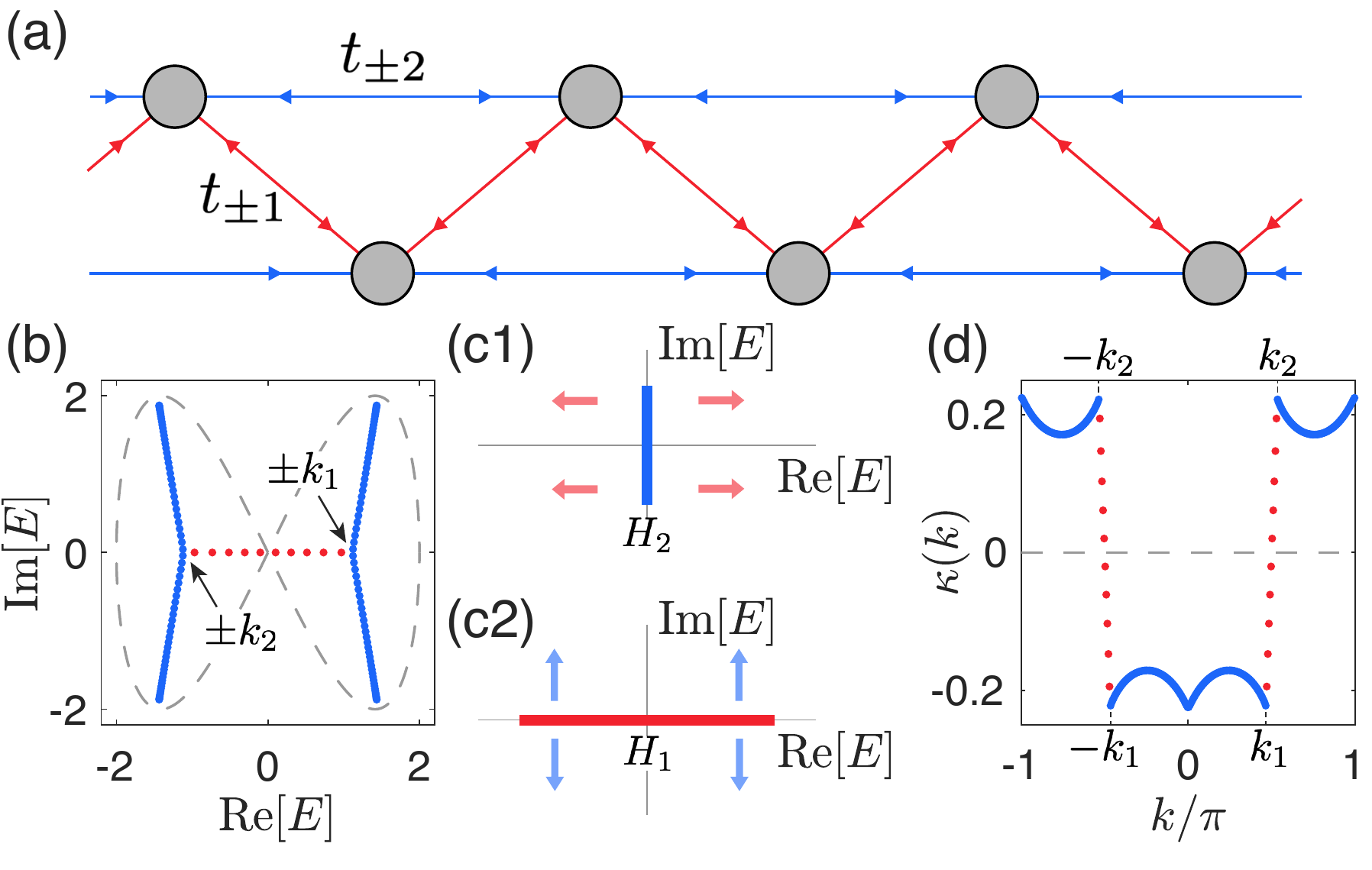}
\caption{
(a) Schematic of our model Eq.~1 with next-nearest couplings.
(b) PBC (gray dash loop) and OBC (blue and red dots) spectra of the system.
(c) Cartoon representation of how the OBC spectrum is affected (red arrows) by adding $H_1$ on $H_2$'s spectrum (blue line), and vice versa.
As discussed in the main text, while $H_1$ and $H_2$ both ``stretch" the OBC spectrum along real and imaginary axis respectively, a pseudo-gap in real energy with sparse eigenmodes ($10$ dots in this example) emerges out of asymmetry.
(d) The inverse decaying length $\kappa(k)$ as $k$ varies from $-\pi$ to $\pi$. Blue and red dots correspond to the eigenenergies in (b).
Parameters in (c) and (d) are $t_1=t_{-1}=t_2=-t_{-2}=1$, with $L=150$ sites. 
}
\label{fig:sketches}
\end{figure}

\noindent{\it Minimal model for a NH pseudo-gap.--} 
For concrete illustration of a NH pseudo-gap, we discuss the minimal model Hamiltonian
$H=H_1+H_2$ with
\begin{eqnarray}
H_1&=& \sum_x t_1\hat{c}^\dagger_{x}\hat{c}_{x+1}+ t_{-1}\hat{c}^\dagger_{x+1}\hat{c}_{x},\nonumber\\
H_2&=&\sum_x  t_2\hat{c}^\dagger_{x}\hat{c}_{x+2}+ t_{-2}\hat{c}^\dagger_{x+2}\hat{c}_{x},\label{eq:H_NNN}
\end{eqnarray}
which is the juxtaposition of two Hatano-Nelson (HN) chains~\cite{HN1996prl,HN1997prb,okuma2021non}
$H_1$ and $H_2$ with nearest and next-nearest neighbor couplings [Fig.~\ref{fig:sketches}(a)]. The HN chain is the simplest model with exhibiting NHSE due to nontrivial PBC point-gap~\cite{zhang2019correspondence,okuma2020topological}, with OBC eigenenergies occupying the line segment between $\pm 2\sqrt{t_jt_{-j}}$, $j=1,2$. 
We specialize to real $t_1=t_{-1}$ and $t_2=-t_{-2}$, such that the two HN chains separately possess pure real and imaginary OBC spectra. In momentum space, 
$H(k)=H_1(k)+H_2(k)$,
\begin{eqnarray}
H_1(k)=2t_1\cos k,\, H_2(k)=2i t_2\sin 2k.\label{eq:hk}
\end{eqnarray}
which also gives the PBC spectrum for $k\in [0,2\pi)$.

\begin{figure}
\includegraphics[width=1\linewidth]{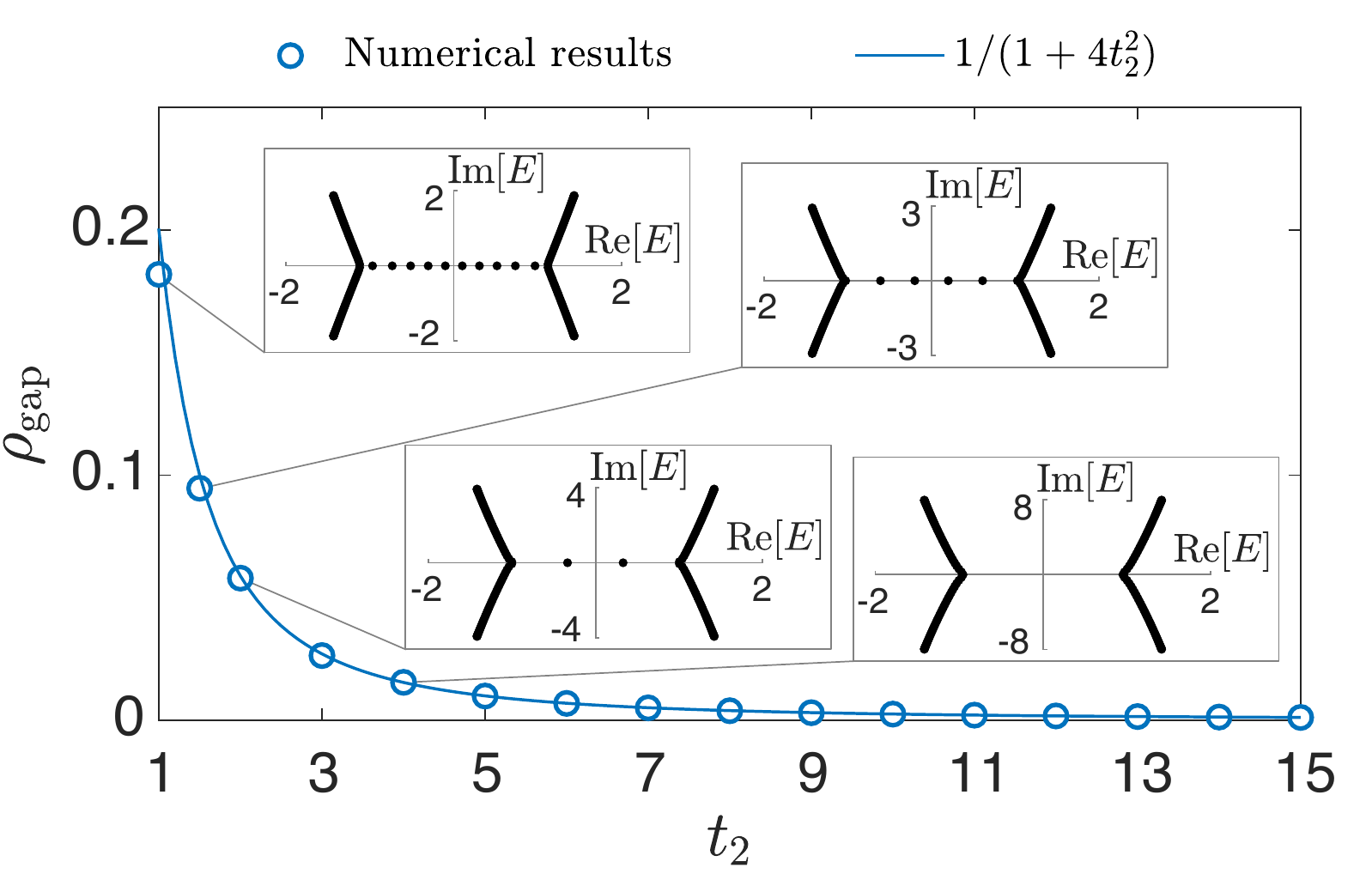}
\caption{The DOS $\rho_{\rm gap}$ within a pseudo-gap as a function of $t_2$. Blue circles are obtained from $2\Delta k/E_{\rm gap}$ numerically, and agrees excellently with the blue line given by Eq.~\ref{eq:rho}.
Insets show the pseudo-gap spectra with $t_1$ and $t_2=1,1.5,2,4$ respectively, obtained with $L=150$ lattice sites. }
\label{fig:delta_k}
\end{figure}

As shown in Fig.~\ref{fig:sketches}(b), the OBC spectrum of $H$ consists of two distinct types of segments: (i) blue arcs at positive and negative $\text{Re}[E]$ with high DOS, separated by (ii) a red segment in the middle with much lower DOS. This middle red segment is dubbed as the NH pseudo-gap because its DOS can be made arbitrarily low by tuning $t_2/t_1$, as further shown in Fig.~\ref{fig:delta_k}, unlike the blue arcs which behave like ordinary bands. 

This NH pseudo-gap is real, and originates from the competition between the NHSE channels of $H_1$ and $H_2$ i.e. 
their differing tendencies to boundary-localize. For our model, this can be traced to the relative sign between $t_1/t_{-1}=1$ and $t_2/t_{-2}=-1$ hopping ratios, which leads to an enigmatic form of competitive interference between boundary localized skin eigenstates. This is most transparently seen on the complex energy plane. Under OBCs, $H_1$ has a purely real spectrum, while $H_2$ has a purely imaginary spectrum. As such, the OBC spectrum of their sum $H=H_1+H_2$ shall be neither purely real or purely imaginary, but a competitive combination of the two. In terms of PBC-OBC spectral flow sketched in Fig.~1, we have a juxtaposition of two spectral flows (of $H_1$ and $H_2$) resembling the left figure, with one set of PBC ellipse plus OBC line segment relatively rotated by $\pi/2$. This is analyzed in careful detail in~\cite{suppmat}. The main idea, as illustrated in Fig.~\ref{fig:sketches}(c), is that $H_1$ can be understood as ``pulling'' the vertical OBC spectrum (blue) of $H_2$ apart, while $H_2$ simultaneously ``pulls'' the horizontal OBC spectrum (red) of $H_1$ vertically. However, due to the dissimilarity between the periodicities of $H_1(k)$ and $H_2(k)$ [Eq.~2], the net effect is that most of the states are ``pulled'' away from the real line into almost vertical OBC arcs (blue), resulting in $H=H_1+H_2$ having a real pseudo-gap of very low DOS [Fig.~\ref{fig:sketches}(b)]. More generally, NH pseudo-gap regions are not necessarily real or straight, but they all originate from analogous competitive mechanisms between different NHSE channels. 

Since pseudo-gaps contain so few states, they appear as steep jumps of the skin inverse decay length $\kappa(k)$ [Fig.~\ref{fig:sketches}(d)]. The values of $\kappa(k)$ before and after the jumps are that of the two pseudo-bands, and can be obtained through the Generalized Brillouin zone (GBZ) \cite{yao2018edge,yokomizo2019non,Lee2019anatomy,lee2020unraveling} (detailed in the Supplemental Materials \cite{suppmat}).
The jumps of $\kappa(k)$ occur as an opposite pair that, for this model, crosses 
$\kappa=0$ twice. In other words, a pseudo-gap between bands that are oppositely localized (with different signs of $\kappa$) may contain delocalized states.

The DOS within the pseudo-gap can be estimated via  $\rho_{\rm gap}=2\Delta k/ E_{\rm gap}$ with $\Delta k=|k_2-k_1|$ the GBZ momentum jump across the pseudo-gap, and $E_{\rm gap}$ the gap size, as shown in Fig.~\ref{fig:delta_k}. As further derived in~\cite{suppmat}, 
\begin{eqnarray}
\rho_{\rm gap}=\sum_{\pm}\frac{d k}{d E}\approx\frac{t_1}{t_1^2+4t_2^2}\label{eq:rho}
\end{eqnarray}
with $\pm$ indexing the jump regions, agreeing closely with numerics [Fig.~\ref{fig:delta_k}]. For $t_1=1$ and $t_2\sim \mathcal{O}(1)$, we can easily obtain $ \rho_{\rm gap}\sim \mathcal{O}(10^{-1})$, much smaller than $\rho\sim \mathcal{O}(1)$ within typical bands. Indeed, even for a larger number of sites ($L=150$), the discrete number of in-gap states already falls to zero at $t_2\geq 4$ (Fig.~\ref{fig:delta_k}).

\begin{figure}
\includegraphics[width=1\linewidth]{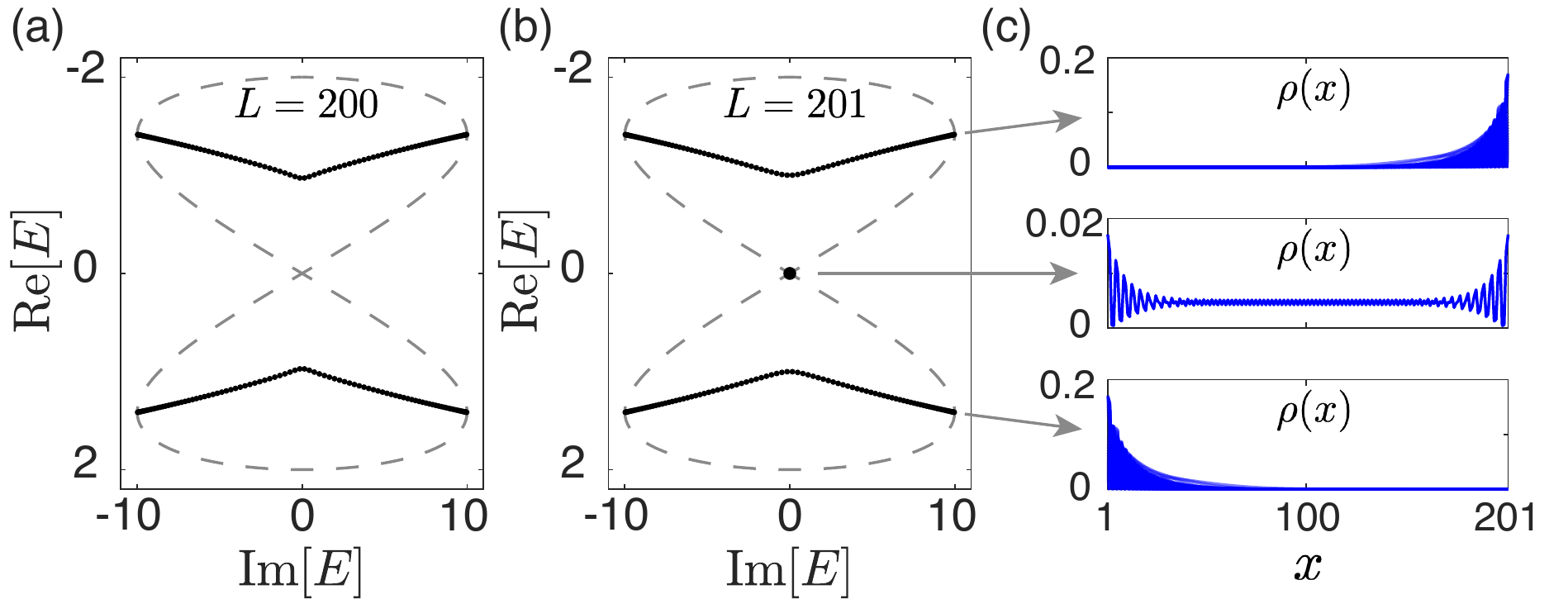}
\caption{(a) and (b) PBC (gray dashed loops) and OBC (black) spectra with $L=200$ and $L=201$ lattice sites respectively, showing a mid-pseudogap mode for odd $L$. (c) Squared wavefunction amplitude $\rho(x)=|\psi(x)|^2$ plots, showing the extended in-gap eigenmode and NHSE-localized pseudo-band eigenmodes.
}
\label{fig:ingap_mode}
\end{figure} 

\noindent{\it Robust extended mid-pseudogap modes. --} 
Like ordinary topological gaps~\cite{shanahan2006atiyah,hasan2010colloquium,essin2011bulk,qi2013axion,gu2016holographic}, NH pseudo-gaps can also host symmetry-protected mid-gap zero modes. However, unlike ordinary in-gaps states which by definition are not extended bulk states, states within NH pseudo-gaps can be extended throughout the sample, and may in fact be \emph{obliged} to be delocalized due to symmetry protection. This scenario represents a ``role-reversal'' from ordinary settings, with \emph{delocalized} mid-pseudogap modes accompanied by \emph{localized} NHSE-deformed bulk skin modes.

\begin{figure*}
\includegraphics[width=1\linewidth]{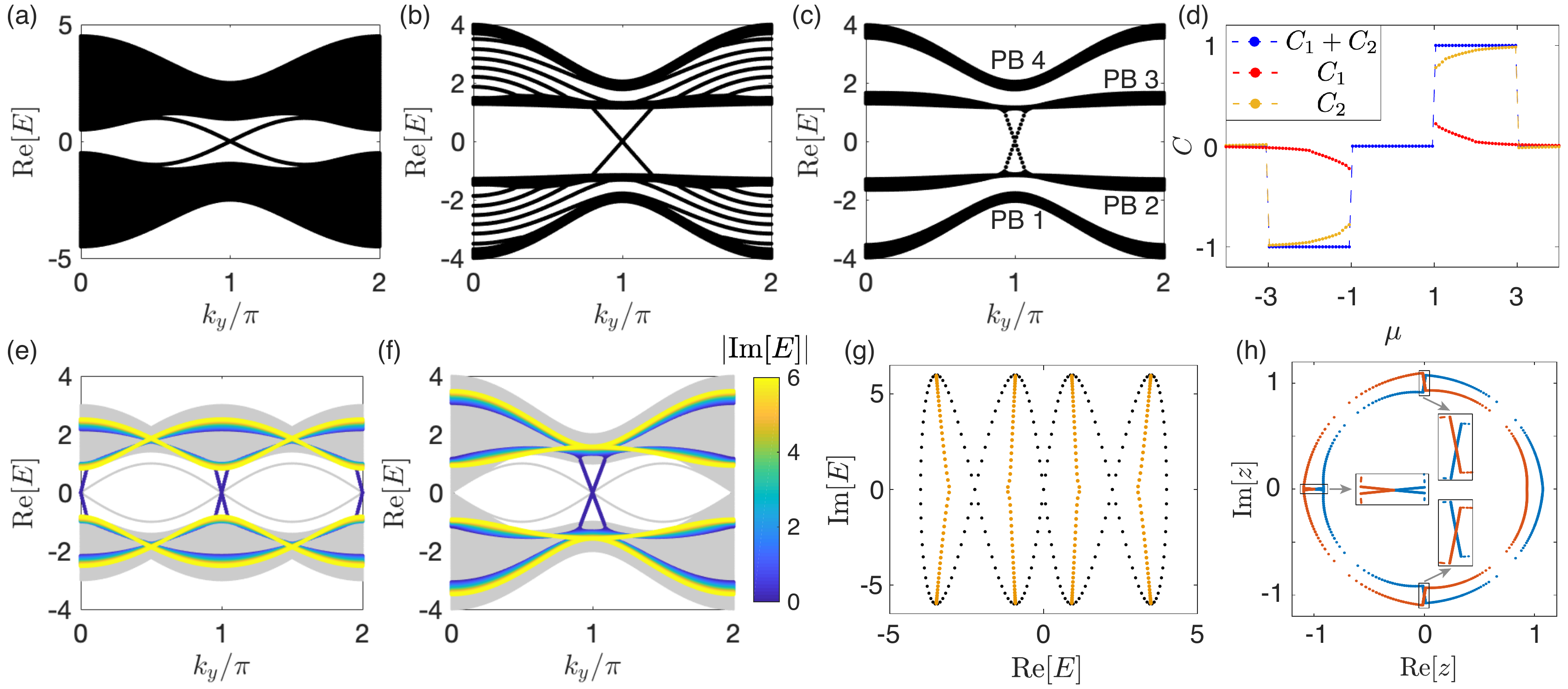}
\caption{
(a-c) Real part of the $x$-OBC spectra for $h_\text{2D}$ in the topological regime ($\mu=1.5$), showing the splitting of the two ordinary bands into four pseudo-bands (PB1 to PB4) as $t_2$ increases from 0 (a) to 1 (b) to 3 (c).
(d) Non-quantized Chern numbers $C_1$ and $C_2$ for PB1 and PB2 of the case in (c), computed with $L=20$ $x$-unit cells over the region where they are pseudo-gapped and well-defined. Their sum $C_1+C_2$ is quantized just like the lower ordinary band. 
(e-f) Real part of the $x$-OBC spectra of $h_{\rm 2D}$ in the trivial regime $\mu=0$ (e) and transition point $\mu=1$ (f). Gray curves represent the Hermitian case with $t_2=0$, and colored curves represent the full non-Hermitian case with $t_2=3$. In panel (f), the pseudo-gap at $\text{Re}[E]=0$ separates upper and lower bands and is traversed by chiral edge modes, just like an ordinary gap, despite being at the topological transition point. 
(g) PBC (black) and $x$-OBC (orange) spectra for the system in (f) at $k_y=0$, with four pseudo-bands separated by three pseudo-gaps.
(h) 1D GBZ solutions $z=e^{ik_x}e^{-\kappa_x}$ of the case in (g) with red(blue) representing $\text{Re}[E]>0$($\text{Re}[E]<0$) bands. The three jumps correspond to the three pseudo-gaps.
Other parameters are $t_x =1$, $t_y =t_{ab}^x =t_{ab}^y =0.5$ and $L=100$ unit cells, unless mentioned otherwise.
}
\label{fig:2D}
\end{figure*}

In our model, the DOS within the pseudo-gap decreases rapidly as $t_2$ increases [Fig.~\ref{fig:delta_k}], such that there are no ``bulk'' eigenmode in the pseudo-gap for a finite system with sufficiently large $t_2$, as in Fig.~\ref{fig:ingap_mode}(a) with $L=200$ sites. Nevertheless, a robust in-gap eigenmode emerges at $E=0$ for odd $L$ [Fig.~\ref{fig:ingap_mode}(b)].
Unlike conventional topological in-gap modes, this pseudo-gap mode is extended along the 1D chain [~\ref{fig:ingap_mode}(c)], since its eigenenergy lies within the PBC spectrum and is hence free from the NHSE \cite{zhu2021delocalization}. 
Yet it is also symmetry-protected like conventional 1D topological edge modes, 
since our real-space model Hamiltonian Eq.~\ref{eq:H_NNN} satisfies $SHS^{-1}=-H^\dagger$ 
with $S={\rm diag}\{1,-1,1,-1,...\}$, which represents chiral symmetry for non-Hermitian systems \cite{Kawabata2018nonHclass,li2019geometric}.
Due to this symmetry, for any $H|\Psi\rangle=E|\Psi\rangle$, we have
\begin{eqnarray}
H^{\dagger}S|\Psi\rangle=-ES|\Psi\rangle,
\end{eqnarray}
meaning that $S|\Psi\rangle$ is a left eigenstate of $H$ localized together with the right eigenstate $|\Psi\rangle$. 
Hence there must be another right eigenstate $|\Psi'\rangle$ at eigenenergy $-E^*$, biorthogonal-paired with the left eigenstate $S|\Psi\rangle$ and localized at the opposite boundary \cite{kunst2018biorthogonal}, forming a chiral pair together with $|\Psi\rangle$. 
Therefore, with an odd number of lattice sites, there will be an unpaired eigenmode with balanced distribution in both directions satisfying $|\Psi'\rangle= |\Psi\rangle$ and $E'=-E^*$, implying purely imaginary eigenenergy and lying within the pseudo-gap of the two symmetric pseudo-bands. 
Note that chiral symmetry itself does not guarantee zero energy for this in-gap eigenmode: this also requires  spinless time-reversal symmetry (TRS) $H=H^*$ [or $H(k)=h^*(-k)$], under which a time-reversal pair of eigenstates shall have eigenenergies with opposite imaginary parts, leading to real energy for the self-paired eigenstate (which is hence pinned to zero in this case, since it must also be pure imaginary). 
See \cite{suppmat} on the consequences of breaking this symmetry.

\noindent{\it Pseudo-Chern bands and non-quantized Chern numbers. --} We next consider pseudo-gaps within bands of nontrivial band topology, and find that notions of band topology i.e. the topological bulk-boundary correspondence become murky: (i) a pseudo-gap splits a band into two or more pseudo-bands with non-integer Chern numbers, yet the pseudo-bands are still connected by integer numbers of edge modes; (ii) a pseudo-gap can remain open when the ordinary gap is closed, allowing in-(pseudo)gap chiral edge modes to exist even at the topological transition point or the trivial regime, in defiance of the total Chern number within ``occupied" pseudo-bands. 

We consider a non-Hermitian 2D Hamiltonian $h_{\rm 2D}(\mathbf{k})=h_{\rm QWZ}(\mathbf{k})+h_2(\mathbf{k})\mathbb{I}$, with
\begin{eqnarray}
h_{\rm QWZ}(\mathbf{k})&=&(2t_x\cos k_x+2t_y\cos k_y +\mu)\sigma_z\nonumber\\
&&+2t^{ab}_y\sin k_y\sigma_y+2t^{ab}_x\sin k_x \sigma_x\label{eq:2D_2}
\end{eqnarray}
the Qi-Wu-Zhang (QWZ) model with nontrivial Chern topology \cite{qi2006topological}, and $h_2(\mathbf{k})=2i t_2\sin 2k_x$ the same anti-Hermitian term as in Eq.~1.
Figs.~\ref{fig:2D}(a-c) illustrates the x-OBC spectra of $h_{\rm QWZ}$ as the non-Hermiticity parameter $t_2$ is increased from 0 to 1 and then 3, in the topologically non-trivial regime of $\mu=1.5$. As $t_2$ increases, the two ordinary bands split into four pseudo-bands (PB1 to PB4 in Fig.~\ref{fig:2D}(c)). A pair of chiral edge modes initially connects both bands, but eventually connect only two pseudo-bands when $t_2=3$. Even though the pseudo-bands cannot be distinguished from ordinary bands~\footnote{Under OBCs, band eigenstates are not labeled by momentum, and a pseudo-band represent an isolated cluster of eigenenergies just like an ordinary band.}, they interestingly possess non-integer Chern numbers despite being connected by integer numbers of Chern edge states. This is presented in Fig.~\ref{fig:2D}(d), where $C_1$ and $C_2$ are the Chern numbers of PB1 and PB2 computed from the Berry curvature in the GBZ $(k_x+i\kappa_x(\bold k),k_y)$ i.e. of the Surrogate Hamiltonian~\cite{lee2020unraveling}. The sum $C_1+C_2$ is quantized as expected of the ordinary Chern band, with transitions at $|\mu|=1$, independent of $t_2$ for this model. In the topologically trivial $|\mu|<1$ phase, the Hermitian QWZ model ($t_2=0$) possesses trivial in-gap edge modes with two crossing, shown in gray in Fig.~\ref{fig:2D}(e). With a nonzero $t_2$, the edge modes partially merge into the bulk to become counter-propagating chiral modes \cite{lababidi2014counter,
zhou2014aspects,zhou2014floquet,ho2014topological,fulga2016scattering,hockendorf2019universal,umer2020counter}.

Interestingly, well-defined edge modes can also traverse a pseudo-gap even if the Chern number is not well-defined, such as the $\mu=1$ case at the topological transition. This is because gapless bands can still \emph{look} gapped with a large pseudo-gap. At $t_2=0$, the OBC spectrum is gapless at $k_y=0$. But at nonzero $t_2=3$, a third pseudo-gap at ${\rm Re}[E]=0$ opens up in the x-OBC spectrum [Fig.~\ref{fig:2D}(g)], in addition to the two pseudo-gaps at nonzero $\text{Re}[E]$, even though the PBC spectrum remains gapless at $k_y=0$. Notably, this pseudo-gap is traversed by chiral modes which are not immediately canceled by another counter-propagating pair at $k_y=0$ 
after entering the topologically trivial phase (see ~\cite{suppmat}).
To affirm that these three gap-like regions are indeed pseudo-gaps, we examine its GBZ $z=e^{ik_x}e^{-\kappa_x}$ at $k_y=0$ [Fig.~\ref{fig:2D}(h)], which is seen to display three sharp jumps in the radius $|z|$, reminiscent of the 1D case in Fig.~\ref{fig:sketches}(d)], suggesting three pseudogaps with vanishing DOS. Specifically, the two jumps at ${\rm Re}[z]=0$ occur within ordinary bands (continued by the same color), corresponding to the pseudo-gaps with positive and negative real energies respectively. The third pseudo-gap at ${\rm Im}[z]=0$ connects different ordinary bands (continued by different colors), 
representing the pseudo-gap at $E=0$. 

Chern pseudo-bands harbor interesting implications for both experiment and theory. Despite existing in non-Hermitian systems with complex spectra, they harbor real Chern chiral modes which will be 
convenient for experimental detection in photonic lattices~\cite{ozawa2019topological},
especially with larger group velocities from more pronounced pseudo-gaps. 
 These chiral edge modes can even be well separated from bulk bands at a topological phase transition, where the system is technically gapless, but appearing gapped due to a pseudo-gap.
The lack of integer Chern number quantization of seemingly isolated pseudo-bands will also lead to renewed theoretical discourse on fractional Chern phases. Despite being supposedly 
lattice analogs of fractional quantum Hall phases~\cite{sun2011nearly,neupert2011fractional,lee2014lattice,anisimovas2015role,neupert2015fractional,lee2017band,kourtis2017weyl}, the requirement of accordingly fractionally occupied Chern bands have been questioned~\cite{kourtis2014fractional,huang2018invariance}, and our discovery of isolated pseudo-bands with non-integer Chern numbers brings about an additional avenue of intrigue. 

\noindent{\it Discussion. --}
We have uncovered a new mechanism for creating pseudo-gaps. It is based on a unique form of interference between different NHSE channels, and arise generically when the complex PBC spectrum geometrically resembles a conductor surface that gives rise to divergent electric fields. Despite having entanglement entropy with gapless logarithmic scaling~\cite{suppmat}, NH pseudo-gaps host enigmatic states like delocalized in-gap modes, as well as Chiral edge modes that seem to defy the topological bulk-boundary correspondence.

The notion of NH pseudo-gaps spurs various new fruitful directions. 
Robust eigenmodes within pseudo-gaps may be exploited for sensing applications since they are extended yet significantly modified by symmetry-breaking terms. Finally, the decomposition of topological bands into isolated pseudo-bands may post profound implications for quasi-particles fractionalization.

\noindent{\it Acknowledgements.--}
 L. L. acknowledges funding support by the Guangdong Basic and Applied Basic Research Foundation (No. 2020A1515110773).

\bigskip

\clearpage

\onecolumngrid
\begin{center}
\textbf{\large Supplementary Materials}\end{center}
\setcounter{equation}{0}
\setcounter{figure}{0}
\renewcommand{\theequation}{S\arabic{equation}}
\renewcommand{\thefigure}{S\arabic{figure}}
\renewcommand{\cite}[1]{\citep{#1}}

\section{Pseudo-gap from the competition/interference between $H_1(k)$ and $H_2(k)$ Hatano-Nelson models} 
In this section ,we give a detailed discussion about how the pseudo-gap emerge from the competition between $H_1(k)=2t_1\cos k$ and $H_2(k)=2it_2\sin 2k$, for the model of Eq. 1 in the main text.
Following the theory of the generalized (non-Bloch) Brillouin zone for systems exhibiting the NHSE~\cite{yao2018edge,Lee2019anatomy,lee2020unraveling}, the OBC spectrum must form lines or curves enclosing zero area in the complex plane, so that pairs of eigenenergies with different $k$ merge into the same point on the energy plane i.e. are degenerate in energy. Such a condition is satisfied by each of $H_1(k)$ and $H_2(k)$, 
with $H_1(k)=H_1(-k)$ and $H_2(k_0-k)=H_2(k_0+k)$ respectively, where $k_0=\pm\pi/4$ or $\pm3\pi/4$. However, the overall system does not satisfy either of them, and its fate depends on the competition between these merging effects from $H_1(k)$ and $H_2(k)$ respectively.

To give a qualitative glance, we first consider a pair of momenta $k_\pm=\pi/4\pm\Delta k$ with $\Delta\CH{k}\in[0,\pi/4]$, where the eigenenergies merge into the same point and form a line-spectrum along the imaginary axis in the absence of $H_1(k)$ (e.g. see Fig. 2(c1) in the main text).
A nonzero $t_1$ separates these two points by assigning different real energies $E_1(k_\pm)=2t_1\cos k_\pm$ to the total energies $E(k_\pm)$.
To cancel out this separation, a complex deformation of the momentum $k_\pm\rightarrow k_\pm+i\kappa$ is required, 
and the eigenenergies of the overall system satisfy
\begin{eqnarray}
E&&(k_+ +i\kappa)-E(k_-+i\kappa)=\nonumber\\
&&-\sin \Delta k[\sqrt{2}t_1+4t_2\cos\Delta k(e^{-\kappa}-e^{\kappa})](e^{-\kappa}+e^{\kappa})\nonumber\\
&&+i\sqrt{2}t_1\sin\Delta k(e^{-\kappa}-e^{\kappa}).
\end{eqnarray}
Here a nonzero $\kappa$ also indicates a non-Hemritian skin localization of the corresponding eigenmode under OBC.
We can see that the real-energy separation between $E(k_\pm)$ induced by $t_1$ is canceled out by a $\kappa$ with
\begin{eqnarray}
e^{\kappa}-e^{-\kappa}=\frac{\sqrt{2}t_1}{4t_2\cos \Delta k}.\label{eq:kappa}
\end{eqnarray}
Note that while $\kappa$ cancels the separation in real energy, it induces another separation in imaginary energy $\delta E_i$ at the same time, meaning that the merging of different $k$ modes does not occur exactly for the pair of $k_\pm$. Nevertheless, it suggests that the two merging points are close to $k_\pm$, especially when $\kappa$ is small and hence the strength of the imaginary separation $\delta E_i$ is weak. 
Thus we can obtain some quantitative conclusions from Eq..~\ref{eq:kappa}:

i). the value of $\kappa$ decreases when $\Delta k$ decreases, meaning that the merging effect from $H_2(k)$ (i.e. between $k_\pm$) dominates with smaller $\Delta k$, i.e. closer to $k=\pi/4$;

ii). the value of $\kappa$ also decreases when $t_2/t_1$ increases, meaning that the system favors the merging effect more for larger $t_2$. 

Combining these conclusions,  we can see that when $t_2$ gets larger, the eigenvalues in a wider range of $\Delta k$ tends to follow the merging effect from $H_2(k)$, or in other words, the skin effect of the corresponding eigenmodes are less affected by the presence of $H_1(k)$.
On the other hand, even when $t_2\gg t_1$, the real part of $E(k_\pm + i\kappa)$ cannot be zero. This is because both of $E(k_\pm)$ have positive real energies, and $E(k_\pm + i\kappa)$ is obtained by erasing the difference between them, i.e. 
\begin{eqnarray}
{\rm Re}[E(k_\pm + i\kappa)]\propto {\rm Re}[E(k_+)+E(k_-)]/2\neq0.
\end{eqnarray}
As a matter of fact, we shall have $\kappa\rightarrow 0$ when $t_2\gg t_1$, and ${\rm Re}[E(k_\pm + i\kappa)]$ ranges from $t_1$ (when $\Delta k=\pi/4$) to $\sqrt{2}t_1$ (when $\Delta k=0$).
Finally, similar results can be obtained for pairs of momenta around $k=-\pi/4,3\pi/4,-3\pi/4$ respectively, with the latter two give negative real energies ranging from $-t_1$ and $-\sqrt{2}t_1$ for them. In short, as sketched in Fig.~2(c1) in the main text, different parts of the imaginary spectrum of $H_2(k)$ tend to acquire positive and negative real values due to a nonzero $H_1(k)$, inducing a non-Hemritian pseudo-gap between $t_1$ and $-t_1$ for the whole system.

Following the same spirit, we may also check another pair of momenta $k'_\pm=\pm(\pi/2-\Delta k')$, where the eigenenergies merge into the same point and 
give a real spectrum
in the absence of $H_2(k)$ (e.g. see Fig. 2(c2) in the main text). Similarly, with a nonzero $t_2$, a separation in their imaginary energies emerges between the eigenmodes of $k'_\pm$, which can also be erased through a complex deformation $i\kappa'$ of the momenta. Namely we have
\begin{eqnarray}
&&E(k'_+ +i\kappa')-E(k'_-+i\kappa')=\nonumber\\
&&+i\cos \Delta k' [t_1(e^{-\kappa'}-e^{\kappa'})+2t_2\sin\Delta k'(e^{-2\kappa'}+e^{2\kappa'})],\nonumber\\
\end{eqnarray}
and the imaginary separation is erased when
\begin{eqnarray}
\frac{t_1}{2t_2\sin\Delta k'}=e^{\kappa'}-e^{-\kappa'}+\frac{2}{e^{\kappa'}-e^{-\kappa'}}.\label{eq:kappa_prime}
\end{eqnarray}
Thus we obtain the following conclusions:
(i) At $\Delta k'=0$, the two points of $k'=\pm\pi/2$ always merge into each other at $\kappa'=0$, meaning that they always obey the merging effect from $H_1(k)$ and exhibit no skin localization.
(ii) For $\Delta k'\neq0$, a nonzero $\kappa'$ is required to erase the imaginary separation of the eigenenergy. The value of $\kappa'$ increases with $\Delta_k'$ before it reaches $e^{\kappa'}-e^{-\kappa'}=\sqrt{2}$, where the right-hand side of Eq.~\ref{eq:kappa_prime} reaches its minimum. 
(iii) The system tends to follow the merging effect from $H_1(k)$ when $t_1\gg t_2$. Here $E(k'_\pm + i\kappa')$ is obtained by erasing the difference between ${\rm Im}[E(k'_\pm)]$, leading to 
\begin{eqnarray}
{\rm Im}[E(k'_\pm + i\kappa')]\propto {\rm Im} [E(k'_+)+E(k'_-)]/2=0.
\end{eqnarray} 
Therefore
 no (imaginary) pseudo-gap emerges in this case, and the eigenenergies are all real when the merging effect of $H_1(k)$ dominates.

Combining the above two scenario, the system at a given momentum shall obey the merging effect from either $H_1(k)$ or $H_2(k)$, depending on which one comes with a smaller complex deformation ($\kappa$ or $\kappa'$) of the momentum of concern. And a real pseudo-gap emerges when $t_2$ is large enough, where a majority of the system obeys the merging effect of $H_2(k)$. 
This can be seen in Fig. 2(b) in the main text, where the pseudo-gap structure is already clear only ten eigenmodes fall with the regime between $E=\pm t_1$ in a system with $L=150$ sites.

In more generic models, an analogous analysis that reveals similar asymmetry arising from the competition/interference of the NHSE from different subsystems will result in a pseudo-gap region of low DOS.

\section{Detailed computation of low density of states}
\subsection{Formalism}
We next explicitly compute the density of states within the pseudo-gap of our 1D model (Eq.~1 of the main text). To do so, we first provide some general results for computing the DOS along an interval with real eigenvalues (which is the case of our pseudo-gap).

Consider a PBC Hamiltonian with dispersion of the form
\begin{equation}
E(z)=\sum_j a_jz^j
\end{equation}
where $z=e^{ik}$ and the $a_j$s are all real. When $|a_j|\neq|a_{-j}|$, $H(z)$ generically traces out a PBC loop, and undergoes the the non-Hermitian skin effect (NHSE) under OBCs. Suppose that there exist a certain segment of the OBC spectrum which is real i.e. $\text{Im}[E(e^{i\tilde k})]=0$ for $k\in [k_\text{min},k_\text{max}]$, where $\tilde k = k + i \kappa(k)$. The objective of this subsection is to solve for the OBC DOS along this real line in energy space. We remind the reader that the results obtained here are only valid if the actual OBC eigenenergies involved are indeed real; the purpose here is not to check whether this assumption is true, but to determine the OBC DOS should the eigenenergies lie on the real line.

We write $z=re^{ik}$ where $r=e^{-\kappa(k)}$. For $k\in [k_\text{min},k_\text{max}]$, 
\begin{equation}
\text{Im}[E(re^{ik})]=\sum_j a_jr^j\sin jk = 0
\label{Im}
\end{equation}
and 
\begin{equation}
\epsilon=\text{Re}[E(re^{ik})]=\sum_j a_jr^j\cos jk 
\end{equation}
where $\epsilon$ refers to the real eigenenergy. The DOS $\rho = \frac{dk}{d\epsilon}$ can be solved in terms of $k$ if $r=e^{-\kappa(k)}$ can be explicitly expressed in terms of $k$ i.e. if $\kappa(k)$ can be explicitly solved. To do, one expresses all the $\sin jk$ in Eq.~\ref{Im} in terms of polynomials in $\sin k$, such that Eq.~\ref{Im} becomes a bivariate polynomial in terms of $r$ and $\sin k$. If $r$ can be solved for, we will be able to obtain a closed form solution to the DOS.


\subsection{Specialization to our 1D example}
We consider the 1D model discussed in Eq.~1 of the main text, whose energy dispersion read (with real $t_1,t_2$)
\begin{eqnarray}
E(z) = 2t_1\cos k + 4it_2\cos k \sin k = t_1z+\frac{t_1}{z}+t_2z^2-\frac{t_2}{z^2}.
\end{eqnarray}
We have 
\begin{equation}
0=\text{Im}[E(re^{ik})]=t_1\left(r-\frac1{r}\right)\sin k + t_2\left(r^2+\frac1{r^2}\right)\sin 2k
\label{Im2}
\end{equation}
\begin{equation}
\epsilon=\text{Re}[E(re^{ik})]=t_1\left(r+\frac1{r}\right)\cos k + t_2\left(r^2-\frac1{r^2}\right)\cos 2k
\label{Re2}
\end{equation}
such that Eq.~\ref{Im2} simplifies to
\begin{equation}
\cos k = -\frac{t_1(r-\frac1{r})}{2t_2\left(r^2+\frac1{r^2}\right)}=\frac{t_1\sinh \kappa}{2t_2\cosh 2 \kappa}.
\label{Im3}
\end{equation}
where $\kappa=\kappa(k)$. Substituting Eq.~\ref{Im3} into Eq.~\ref{Re2}, we obtain
\begin{equation}
\epsilon=2t_2\sinh 2\kappa + \frac{t_1^2\sinh 2\kappa}{2t_2 \cosh^2 2\kappa}.
\label{Re3}
\end{equation}
To proceed, we express $\epsilon$ in terms of $k$ by inverting Eq.~\ref{Im3} to obtain $\kappa$ in terms of $k$: 
\begin{equation}
\sinh \kappa = \frac1{8t_2\cos k }\left[t_1-\sqrt{t_1^2-32t_2^2\cos^2k}\right].
\end{equation}
The negative solution branch has been chosen as the smaller $\kappa$ solution dominates the larger $\kappa$ solution viz. $~e^{-\kappa x}$. Substituting it into Eq.~\ref{Re3}, we obtain the DOS $\rho$ via
\begin{equation}
\rho = \sum\left|\frac{d\epsilon}{dk}\right|^{-1}=\sum\left|\sin k\frac{d\epsilon}{d(\cos k)}\right|^{-1}
\label{rho}
\end{equation}
where we have included a sum over all the different $k$ that contribute to each OBC eigenenergy (at least two unique $k$ solutions are required for OBC spectra, which has to satisfy the boundary conditions on both sides ). So far, no approximation has been made.

\subsubsection{Smallness of the density of states}
If we examine Eq.~\ref{Im3} again, we note that the maximal value of $|(t_2/t_1)\cos k| = \left|\frac{\sinh \kappa}{2\cosh 2 \kappa}\right|$ is $\frac1{8}\left(\frac1{\sqrt{2-\sqrt{3}}}-\sqrt{2-\sqrt{3}}\right)\approx 0.177$, which is rather small. Note, however, this does \emph {not} give the window of $k$ that realizes real OBC eigenvalues, because nowhere was the condition $\kappa_1=\kappa_2$ used. However, what it gives is the drastic simplification $\sinh\kappa \approx 2(t_2/t_1)\cos k$, $\sinh 2\kappa \approx 4(t_2/t_1)\cos k$, $\cosh \kappa \approx \cosh 2\kappa \approx 1$ and of course $k\approx \pm\frac{\pi}{2}$ (remember, it is the small window of $k$ that gives the pseudo-gap its small DOS). Putting these into Eq.~\ref{rho}, we obtain the nice simple result in the main text
\begin{equation}
\rho_\text{tot} = \sum\left|\frac{d\epsilon}{dk}\right|^{-1}\approx \sum_\pm\left|\frac{d[(2t_1+8t_2^2/t_1)\cos k]}{dk}\right|^{-1}\approx\frac{t_1}{t_1^2+4t_2^2}.
\label{rho2}
\end{equation}
Even for $t_2/t_1$ slightly greater than one, we already obtain a rather small DOS. Retaining a few more orders of $\cos k$ in our derivation, we can obtain a more precise expression
\begin{eqnarray}
\rho_\text{tot}\approx\frac{1}{t_1}\frac{t_1^4-6t_2^2t_1^2\cos^2k-114 t_2^4\cos^4k}{t_1^4+4t_2^2t_1^2+(224t_2^4+6t_2^2t_1^2-t_1^4/2)\cos^2k}.
\label{rho3}
\end{eqnarray}

\section{Numerical approach to the inverse decay length $\kappa(k)$}
Here we give a short discussion about how we numerically obtain the generalized Brillouin zone (GBZ), i.e. the value of $\kappa(k)$.
First, the PBC Hamiltonian is rewritten as $h(z)$ with $z=e^{ik}$. Then we solve the characteristic equation $h(z)-E=0$ for $z$ with different complex values of $E$, and label the solutions as $z_{-r}, z_{-r+1},...z_l$ (without $z_0$). 
Here $r$ and $l$ are the longest hopping ranges toward the right- and left-hand sides respectively.
A given $E$ belongs to the OBC spectrum as long as $|z_1|=|z_{-1}|$, and this pair of solutions $z_{\pm1}$ give two points of the GBZ, denoted as $z_{\rm GBZ}$. 
The GBZ is obtained by scanning over different complex energies $E$ and collecting all possible $z_{\rm GMZ}$, and the inverse decaying length $\kappa(k)$ is given by $z_{\rm GBZ}=e^{ik}e^{-\kappa(k)}$.

In the 2D cases with nontrivial topology, the Berry curvature is obtained by complex shifting the momentum by $i\kappa(k)$ to obtain the so-called surrogate Hamiltonian~\cite{lee2020unraveling} $H(k_x+i\kappa(k_x),k_y)$, from which the Chern numbers $C_{1,2}$ for the pseudo-bands are obtained by integrating the Berry curvature (a discrete sum in numerical approach) only in the associated partial GBZ, i.e. $k_x\in[-\pi/2,\pi/2]$ or $k_x\in[\pi/2,3\pi/2]$, with $k_y\in [0,2\pi]$. [Also see Fig. \ref{fig:2D_2}(d)-(e) for the correspondence between different parts of the GBZ and the pseudo-bands.]


\section{Entanglement entropy}
\begin{figure}[H]
\begin{center} 
\includegraphics[width=.6\linewidth]{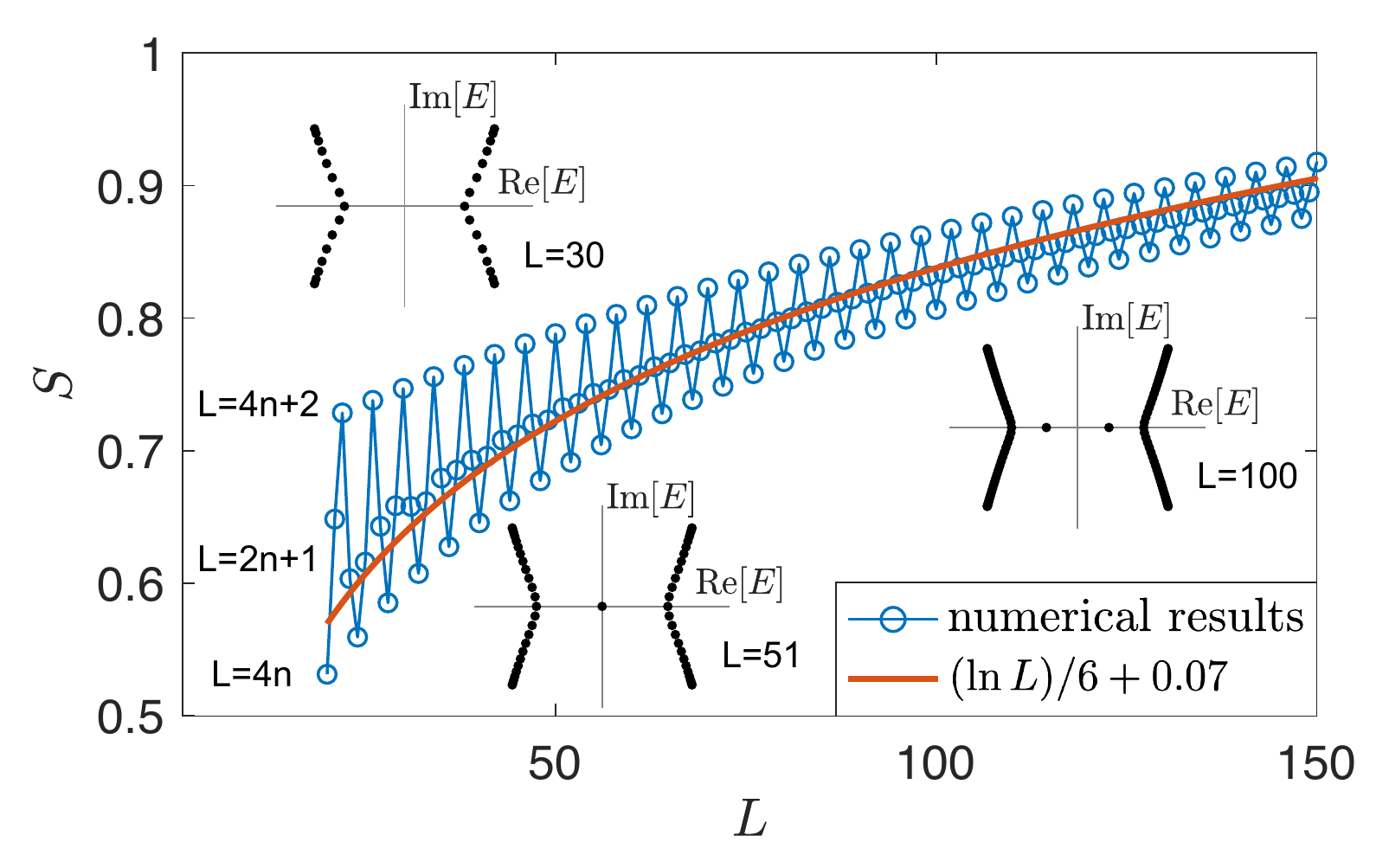}
\end{center} 
\caption{EE versus $L$. $t_1=1$ and $t_2=2$, with $L$ ranging from $20$ to $150$. For odd numbers of $L=2n+1$, the entanglement cut is chosen as the lattice sites with $x\in[(L+1)/2,L]$. Insets illustrate the spectra with $L=30$, $51$, and $100$ respectively.
}
\label{fig:EE}
\end{figure}

In our system, the two parts of the spectrum are separated by a pseudo-gap at finite size, which is filled with sparse eigenmodes when $L$ is large. 
To see that a pseudo-gap is not a genuine gap, we compute its biorthogonal entanglement entropy (EE) for a chosen entanglement cut, as described in~\cite{li2020critical,chang2020entanglement}:
\begin{eqnarray}
S=-\sum_j [\eta_j \ln \eta_j+(1-\eta_j)\ln (1-\eta_j)],
\end{eqnarray}
where $\eta_j$ are the eigenvalues of the correlator matrix
\begin{eqnarray}
C_{xy}=\langle G_L|\hat{c}^\dagger_x\hat{c}_y|G_R\rangle.
\end{eqnarray}
Here $G_L$ ($G_R$) are the left (right) many-body ground state,
and we consider the case of half-filling , i.e. only the single-particle eigenmodes with negative real energies are occupied.
The entanglement cut is chosen as the lattice sites with $x,y\in[L/2+1,L]$ (here we consider an even number of sites only). as the eigenmodes with negative real energies accumulate at the right-hand side.
It is known that the EE obeys an area law scaling for a gapped system, and has a logarithmic dependence on system size for gapless system.
In Fig.~\ref{fig:EE}, we see that the EE obeys a logarithmic scaling $S\approx (\ln L)/6$, even when the spectrum is seen to be gapped. 

Note that EE is obtained solely from the eigenmodes. That is to say, although the eigenenergies of our model give a gapped spectrum, the eigenmodes behave like those of a gapless system.

\section{In-gap extended eigenmode with symmetry breaking}
In the 1D model discussed in the main text, we have shown that the in-gap extended eigenmode is protected by a chiral symmetry, and it is pinned at the zero energy by the spinless time-reversal symmetry (TRS). In this section, we will give more details about how this eigenmode behaves upon symmetry breaking.
\subsection{TRS breaking}
\begin{figure}
\includegraphics[width=1\linewidth]{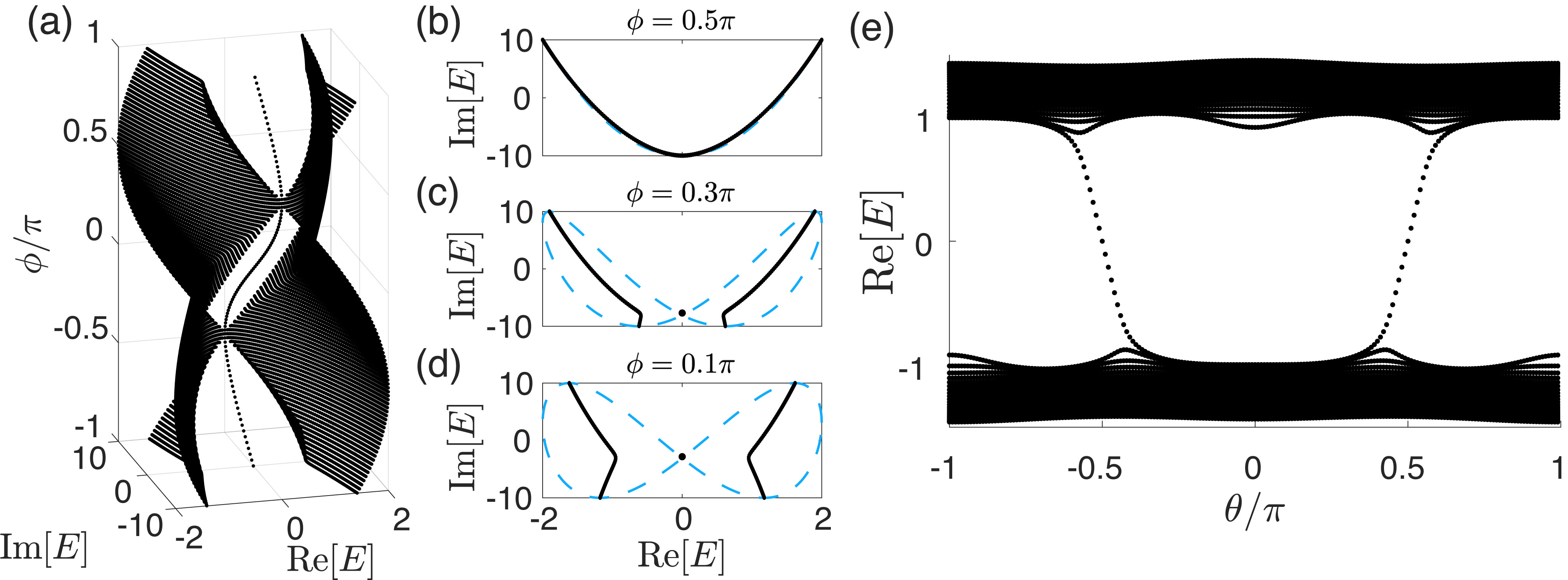}
\caption{(a) OBC spectrum (black dots) versus $\phi$ with $t_{\pm2}=5e^{\pm i\phi}$, and $L=201$. The in-gap eigenmode acquires an imaginary energy when $\phi\neq 0$ or $\pi$. (b-d) PBC (blue dash lines) and OBC (black dots) spectra with different $\phi$ in (a).
The PBC and OBC spectra coincide with each other in (b), with $\phi=0.5\pi$.
(e) Real part of the spectrum of the system with the chiral symmetry broken by $t_2'\cos\theta$. Parameters are $t_2=5$, $t_2'=2$, and $L=101$.
}
\label{fig:spectrum_vs_phi}
\end{figure}

We first keep the chrial symmetry and break the TRS by adding an extra phase $\phi$, i.e. $t_{\pm2}\rightarrow t_2 e^{\pm i\phi}$. As shown in Fig.~\ref{fig:spectrum_vs_phi}, the in-gap eigenmode can move along the imaginary axis with the symmetry unbroken.
On the other hand, this eigenmode always coincides with the PBC spectrum at $k=\pi/2$ [e.g. in Fig.~\ref{fig:spectrum_vs_phi}(b)-(d)], meaning that no complex deformation of $k$ is needed to give raise to it under OBC. Thus the in-gap eigenmode shall has an extended distribution across the system, e.g. as shown Fig. 4(c) in the main text.

\subsection{Chiral symmetry breaking}
Next we keep the TRS and break the chiral symmetry that protect the in-gap eigenmode by adding the following term,
\begin{eqnarray}
H_2'=t_2'\cos\theta\sum_x  \hat{c}^\dagger_x\hat{c}_{x+2}+h.c..
\end{eqnarray}
The single in-gap eigenmode will acquire a real energy and move into the bulk, as shown in Fig.~\ref{fig:spectrum_vs_phi}(e)
The in-gap eigenmode looks like a chiral edge mode, but behaves differently. It exhibits a skin localization toward one direction when its energy is negative, and the other direction when positive. And if we consider a smaller $t_2'$, the in-gap eigenmode does not necessarily goes into the bulk, it may just connect to itself in the pseudo-gap.

\section{Pseudo-bands in a topologically nontrivial 1D system}
Here we consider a topologically nontrivial 1D model, constructed by two copies of the simple 1D chains discussed in the main text with opposite signs, and some extra inter-chain couplings and on-site potentials.
The momentum-space Hamiltonian reads
\begin{eqnarray}
H_{\rm 2-chain}(k)&=&[H(k)+V]\sigma_z+t_{ab}\sin k\sigma_x,\label{eq:hk_topo}\\
H(k)&=&2t_1\cos k+2it_2\sin 2k\nonumber
\end{eqnarray}
with $H(k)$ given by Eq.~\ref{eq:hk}, $V$ a chain-depedent on-site potential, and $t_{ab}$ the inter-chain couplings. 
With $t_2=V=0$, the Hamiltonian vector $\vec{h}(k)=(h_z(k),h_x(k))$ has a nontrivial winding number $\nu_{\vec{h}}$ across the Brillouin zone, leading to the topological properties of this system. Note that this winding is different from the spectral winding number $\nu_E$ that induces the NHSE.
A nonzero $V$ will shift the trajectory of $\vec{h}(k)$, but does not change the topology untill $|V|>|2t_1|$. On the other hand, a nonzero $t_2$ introduce non-Hermiticity to the system, and may induce pseudo-gaps in this system.

\begin{figure}
\includegraphics[width=0.8\linewidth]{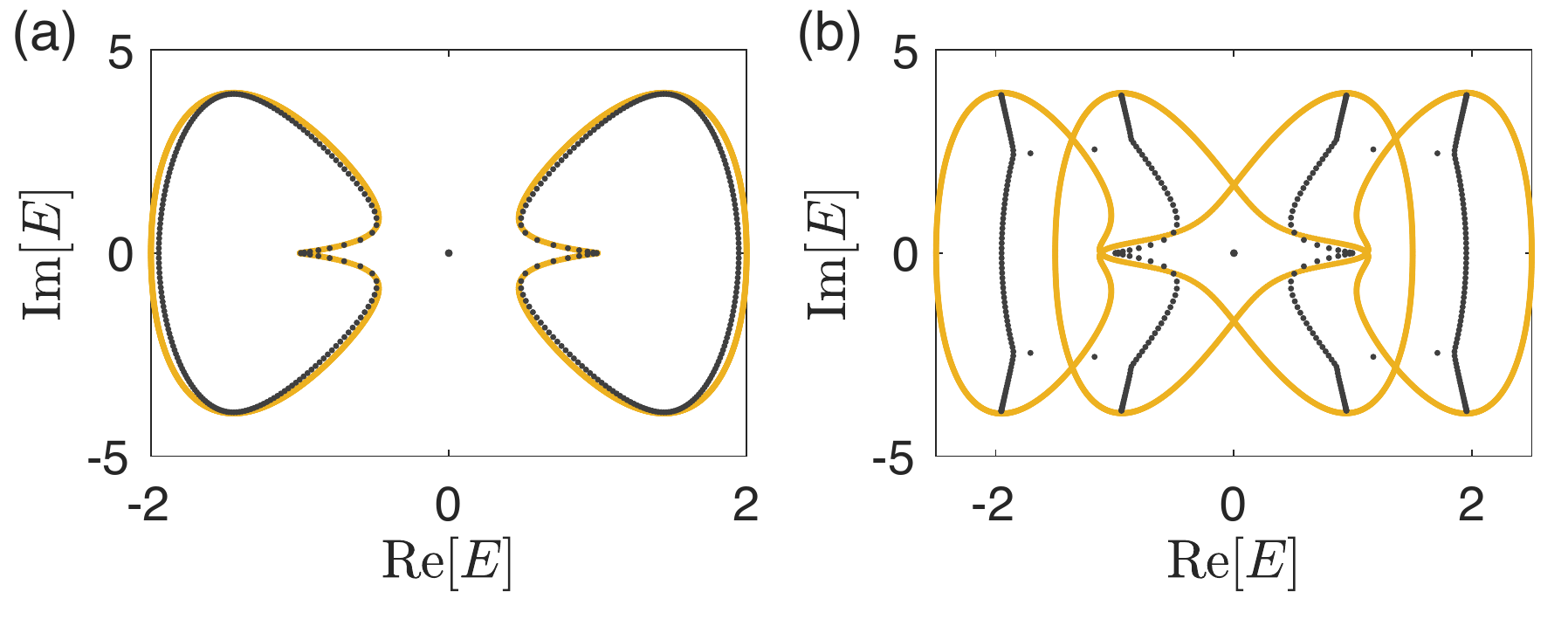}
\caption{PBC (yellow) and OBC (black) spectrum of the system described by Eq.~\ref{eq:hk_topo}, with (a) $V=0$ and (b) $V=0.5$. The black dots at $E=0$ are the topological modes corresponding to the nontrivial winding of $\vec{h}(k)$.
Other parameters are $t_1=1$, $t_2=2$, $t_{ab}=0.5$.}
\label{fig:topo}
\end{figure}

In Fig.~\ref{fig:topo} we illustrate the spectra of two typical cases of this model, where the nontrivial topology gives gives raise to two zero edge modes under OBC (the black dots at $E=0$).
Note that when $V=0$, there is no NHSE under OBC even the PBC spectrum forms a loop. 
This is because the Hamiltonian satisfies the symmetry $\sigma_x H_{\rm 2-chain}(\pi/2+k)\sigma_x=H(\pi/2-k)$ provided $V=0$, 
thus the two segments of the spectrum with $k\in(-\pi/2,\pi/2]$ and $k\in(\pi/2,3\pi/2]$ are identical. On the other hand, the eigenenergies of the system satisfy $E(\pi/2)=E(-\pi/2)$, meaning that each segment is connected to itself head to tail.
Therefore the PBC spectrum winds along the same loop twice with opposite directions (clockwise and anticlockwise), resulting in a zero spectral winding number $\nu_E=0$. The OBC spectrum is also seen to be almost identical with the PBC one. However, with a nonzero $V$, the above mentioned symmetry is broken, 
thus the degeneracy between the two segments is lifted and the NHSE emerges, leading to a divergence between the PBC and OBC spectra. Furthermore, the OBC spectrum now forms four pseudo-bands, with a few isolated in-gap eigenmodes. Meanwhile, the nontrivial topology at $V=0$ is still preserved and the associated zero modes still survive.

The emergence of pseudo-band structure here is actually a bit different from the previous model. 
When $V=t_{ab}=0$, 
the system has two decoupled chains, each has two pseudo-bands as discussed in the main text, denoted here as $E^\pm_1$ and $E^\pm_2$ for the two chains respectively. These two pairs of pseudo-bands are also degenerate to each other, i.e. $E^\pm_1=E^\pm_2$.
A nonzero $t_{ab}$ couples each degenerate pair of the pseudo-bands, e.g. $E^+_1$ and $E^+_2$, and turns them into an energy band of the overall 2-band system.  That is, the gap at $E=0$ of this system (in both panels of Fig. \ref{fig:topo}) is a normal gap, not a pseudo-gap.
On the other hand, a nonzero $V$ lifts this degeneracy, effectively making it more difficult to couple them. Consequently, a pseudo-gap emerges within each of the two normal band in Fig. \ref{fig:topo}(a), giving raise to the four pseudo-bands in (b).

\section{Further results of the 2D model}
\subsection{Finite-size effect of the topological edge modes}
In Fig. 5(f) in the main text, we show that our 2D non-Hermitian system at the topological phase transition point possesses a pseudo-gap, instead of a normal gap closing. Especially, in a finite-size system, the pseudo-gap effectively acts as a normal gap as no eigenmode lies within it, which seems to be contradictory to the notion of gap-protected topology. That is, a pair of chiral edge modes may appear/disappear without different (pseudo-)bands touching each other.

\begin{figure}
\includegraphics[width=1\linewidth]{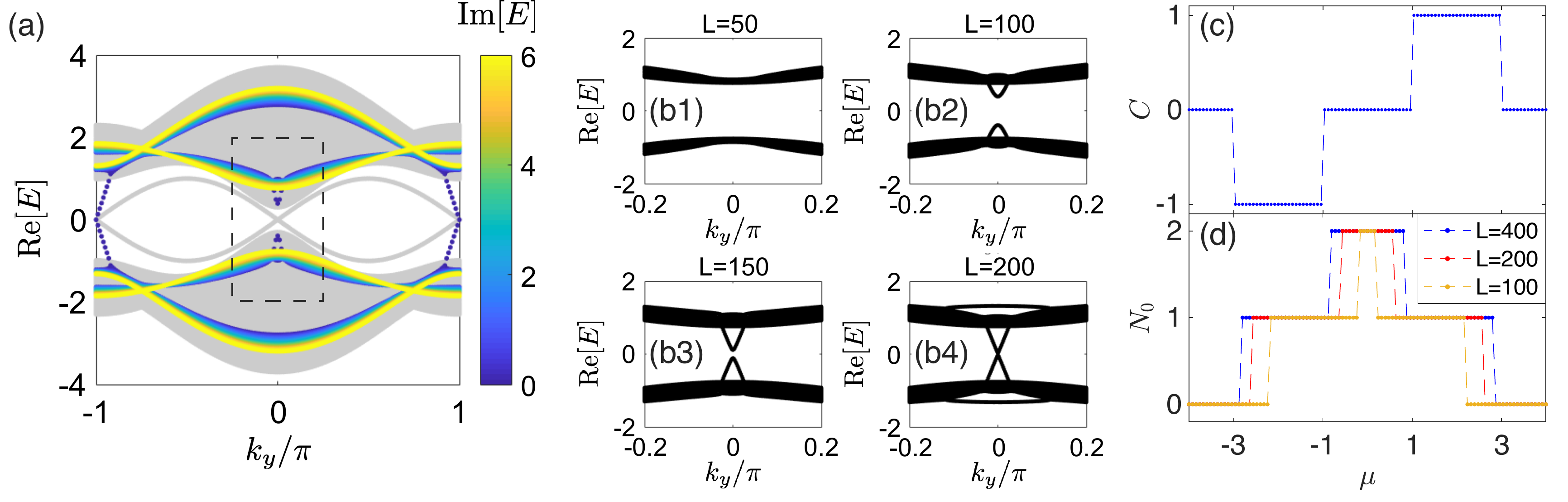}
\caption{(a) The x-OBC spectra with $t_2=0$ (gray) and $t_2=3$ (colored), the system's size is chosen as $L=100$.
(b) The spectrum in the regime enclosed by the black dash line in (a), with different sizes of the system.
(c) Chern number as a function of  $\mu$ of the lower normal band (i.e. the lower two pseudo-bands), obtained  $L=20$ unit cells along $x$ direction. (d) The number of zero energy eigenmodes as a function of $\mu$, with differennt numbers of unit cells along $x$ direction.
Other parameters are $t^x_{ab}=t^y_{ab}=t_y=0.5$, $t_x=1$, and $t_2=3$.}
\label{fig:2D_size}
\end{figure}

In Fig. \ref{fig:2D_size}(a), we illustrate the x-OBC spectrum of the model in a topologically trivial phase with the Chern number $C=0$, where its Hermitian counterpart has a pair of topologically trivial edge modes with two crossing. Yet when the non-Hermiticity is tuned to $t_2=3$, the crossing at $k_y=0$ separates into two gapped branches, while the other one at $k_y=\pi$ evolves into a pair of chiral edge modes. This observation seems to suggest a nontrivial topology with $|C|=1$ in such cases.

To resolve this contradiction, we note that the Chern number describes the topological properties in the thermodynamic limit, but non-Hermitian systems usually suffer significantly from finite-size effect (e.g. in Refs. \cite{budich2020sensor,li2020critical}). In Fig. \ref{fig:2D_size}(b), we zoom in on the regime around $k_y=0$ for the system with different values of $L$, the number of unit cells along $x$ direction. It is seen that with increasing $L$, a pair edge modes emerges from the two pseudo-bands, and connects into a chiral edge modes when $L$ gets large enough. Together with the other pair of chiral edge modes at $k_y=\pi$ in Fig. \ref{fig:2D_size}(a), they form the counter-propagating edge modes and correspond to the zero Chern number.

In Fig. \ref{fig:2D_size}(c) and (d), we compare the Chern number $C$ and the number of pairs of zero-energy modes $N_0$ for different values of $L$.
Here $N_0$ is obtained by numerically counting the zero-energy modes at $k_y=0$ and $k_y=\pi$. For example, the in-gap modes in Fig. \ref{fig:2D_size}(b4) count as $1$, and those in Fig. \ref{fig:2D_size}(b2 and b3) count as $0$.
It is clearly seen that the transition point of $N_0$ matches the topological transition of $C$ better when the size of the system gets larger.

\subsection{Correspondence between the GBZs and the pseudo-bands}
\begin{figure}[H]
\begin{center}
\includegraphics[width=0.8\linewidth]{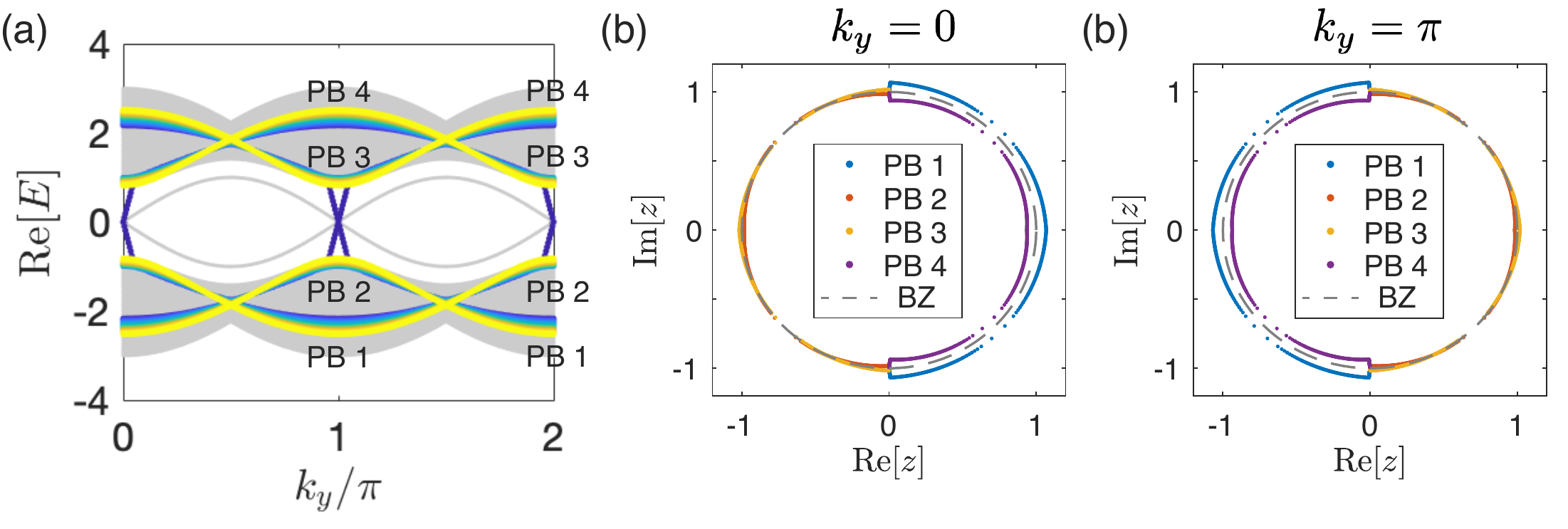}
\includegraphics[width=0.8\linewidth]{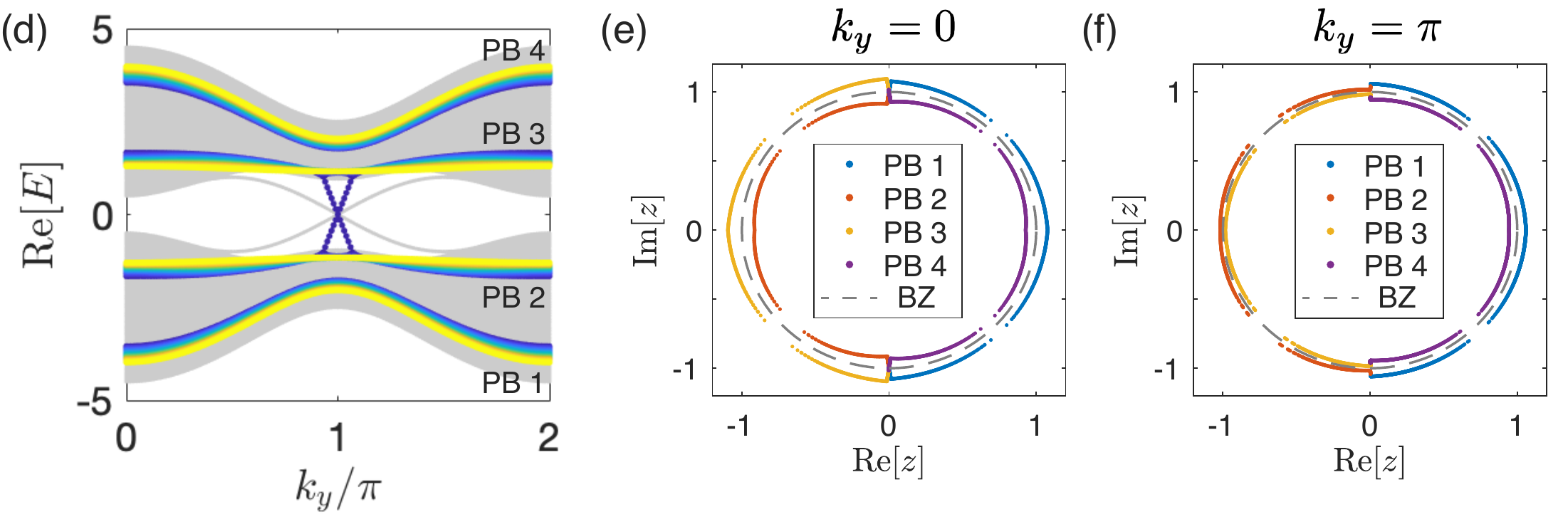}
\end{center}
\caption{(a) The x-OBC spectra with $t_2=0$ (gray) and $t_2=3$ (colored), with four pseudo-bands indicated by PB 1 to PB 4. Note that each pair of pseudo-bands exchanges when $k_y$ varying from $0$ to $\pi$, and from $\pi$ to $2\pi$.
(b) and (c) The GBZs at $k_y=0$ and $\pi$ respectively, colors indicate the correspondence between (parts of) GBZs $z=e^{i k_x}e^{-\kappa_x(k_x)}$ and the pseudo-bands.
Other parameters are $\mu=0$, $t_x =1$, $t_y =t_{ab}^x =t_{ab}^y =0.5$,and $L=100$ unit cells.
(d)-(f) the same as in (a)-(c), but with $\mu=1.5$.
}
\label{fig:2D_2}
\end{figure}

In Fig.~\ref{fig:2D_2}, we further illustrate the correspondence between different parts of the two GBZs and the four pseudo-bands, for two different cases with crossed and fully-separated pseudo-bands, at $k_y=0$ and $k_y=\pi$.


%

\end{document}